\documentclass[twocolumn, showpacs, pre]{revtex4-1}
\usepackage{amssymb,amsfonts,amsmath,amsthm,color,url,bbm, mathrsfs, marvosym}
\usepackage{graphicx}
\usepackage[caption=false]{subfig}
\usepackage{booktabs}
\usepackage[english]{babel}
\usepackage{multirow,tabularx}
\usepackage{hyperref}
\allowdisplaybreaks[1]

\newcommand{\mean}[1]{\left\langle #1 \right\rangle}
\newcommand{\sss}[1]{\scriptscriptstyle{#1}}

\graphicspath{{graphicspdf/}}

\begin{document}
\title{Power-law relations in random networks with communities
}
\author{
Clara Stegehuis
}
\author{
Remco van der Hofstad
}
\author{
Johan S.H. van Leeuwaarden
}
\affiliation{Eindhoven University of Technology, Department of Mathematics and Computer Science, P.O. Box 513, 5600 MB Eindhoven, The Netherlands}


\begin{abstract}
Most random graph models are locally tree-like \--- do not contain short cycles\--- rendering them unfit for modeling networks with a community structure.
We introduce the hierarchical configuration model (HCM), a generalization of the configuration model that includes community structures, while properties such as the size of the giant component, and the size of the giant percolating cluster under bond percolation can still be derived analytically.
Viewing real-world networks as realizations of HCM, we observe two previously undiscovered power-law relations: between the number of edges inside a community and the community sizes, and between the number of edges going out of a community and the community sizes.
We also relate the power-law exponent $\tau$  of the degree distribution with the power-law exponent of the community size distribution $\gamma$. In the case of extremely dense communities (e.g., complete graphs), this relation takes the simple form $\tau=\gamma-1$.
\end{abstract}

\maketitle

\section{Introduction}

Random graphs serve to model large networked systems, but are typically unfit for modeling community structure. Communities refer to relatively densely connected groups of vertices, with more sparse connections between groups, and the community structure refers to an arbitrary number of groups, each of arbitrary size and structure.
In this paper we introduce the Hierarchical Configuration Model (HCM), a model for generating networks as random graphs with not only arbitrary degree distribution but also arbitrary community structure. The HCM is directly applicable to network data, remains solvable in the large-network limit for properties related to component sizes, clustering coefficients and percolation, and is a natural extension of the widely studied configuration model (CM)~\cite{bollobas1980,molloy1995, newman2001} for random graphs with a given degree distribution.

Communities are relatively densely connected and contain relatively many short cycles. Since the CM contains only few short cycles, it cannot model networks with community structure. One possibility to add community structure to random graphs is by adding so-called households~\cite{ball2009,ball2010,coupechoux2014,trapman2007}. In this line of work, on the macroscopic level, the graph is initially a CM in which each vertex of the graph can be replaced by a complete graph (referred to as household). Vertices in a household have links to all other household members, which creates a community structure. These household models allow to study networks with a prescribed degree distribution and a tunable clustering coefficient, because the clustering coefficient can be manipulated by the household structure. Hence, the focus in~\cite{ball2009,ball2010,coupechoux2014,trapman2007} is on locally incorporating short cycles to explain clustering at the global network level. In a similar spirit, a class of random graphs was introduced in~\cite{newman2009} in the form of a random network that only contains random edges and triangles. Each vertex is assigned the number of triangles it is in. The triangles are formed by pairing the nodes at random, and regular edges are formed according to the statistical rules of the CM.
The model in~\cite{newman2009} was extended in~\cite{karrer2010} to networks with arbitrary distributions of subgraphs. 

Like in these previous works, our goal is to develop a more realistic yet tractable random network model, by creating conditions under which the tree-like structure is violated within the communities, but remains to hold at a higher network level \--- the network of communities in our case. There are, however, considerable differences with these earlier works. The model in~\cite{karrer2010,newman2009} departs from a specification of all possible subgraphs or motifs, which is the triangle in~\cite{newman2009} and all possible subgraphs in~\cite{karrer2010}. The network is then created by specifying the number of subgraphs attached to each vertex and then sampling randomly from the set of compatible networks. A community can thus exist of many subgraphs, think of a large cluster of triangles, which makes the framework in~\cite{newman2009,karrer2010} harder to fit on real-world networks. In fact, in~\cite{karrer2010} the appropriate selection of subgraphs and their frequencies for practical purposes is mentioned as a challenging open problem. The approaches in~\cite{ball2009,ball2010,coupechoux2014, trapman2007,newman2009} are geared towards increasing clustering and fitting a global clustering coefficient, but are less suitable to directly describe community structure. Like~\cite{coupechoux2014,trapman2007} we construct a random graph model that at the higher level is a tree-like configuration model, and at the lower level contains subgraphs, but these subgraphs do not need to be complete graphs. Large real-world communities are relatively dense, but not necessarily  {\it completely} connected. We thus generalize the setting of~\cite{ball2009,coupechoux2014,trapman2007} to arbitrary community structures, to account for heterogeneity in  size and internal connectivity.

The HCM breaks away from previous models with clustering or communities, because the model can use any proposed community structure as input. That is, the HCM viewed as an algorithm first models the community structure, and only then creates the random network model.
This top-down approach is in sharp contrast with the bottom-up approach taken in \cite{ball2009,ball2010,coupechoux2014, trapman2007,newman2009}. To be more specific, the community structure can be uncovered by some detection algorithm that, when applied to a real-world network, leads to a collection of plausible communities and their frequencies. By sampling from this collection of communities, the HCM can generate resampled networks with similar structure as the original network. The HCM thus enriches standard random graph models with the ability to describe random yet realistic community structures. Where the CM is the canonical model for complex networks with power-law degree sequences, the HCM adds to this the community structure.


The main contribution of this paper is the introduction and analysis of the HCM. As is common for the CM~\cite{newman2001}, we perform
our study under the assumption of locally tree-like approximations, ignoring the presence of double edges and cycles. Using this {\it tree ansatz} we can apply the generating function formalism~\cite{newman2001} to obtain analytical results. A fully rigorous mathematical treatment of the HCM, taking multiple edges and self-loops into account, is performed in a companion paper~\cite{hofstad2015}. Based on our analysis of the HCM, we discuss several phenomena that each will spark off future research directions.

Our work reveals a potentially crucial property of real-world networks that has received virtually no attention: the joint distribution $p_{k,s}$ of the community size $s$ and the number of connections $k$ a community has to other communities. The size of the giant component delicately depends on this joint distribution, which can be determined from network data once the community structure is determined. In fact, after studying $p_{k,s}$ for several real-world networks, we observe \textit{a power-law relation between the size of the communities and the number of edges out of a community in many real-world networks}.

Except for this joint distribution, the size of the giant component does not depend on detailed information about the structure of the communities. When we perform percolation on the network, more precise structural information does matter. To see this, imagine a process spreading over the network, by starting at some vertex and traversing according to some rule to all vertices in the connected component to which this vertex belongs. Before percolation, once the process reaches a vertex in a community, it reaches the entire community. But after percolation, the community no longer needs to be connected, so that parts of the community may become unreachable. Despite this difficulty, we are able to describe the percolation phase transition explicitly. Inspired by this need to include detailed community structure, and thus extend the model description beyond $p_{k,s}$, we observe a second \textit{power-law relation between the denseness of a community and its size} in several real-world networks.

For the present paper, the most important application of the HCM is to investigate power-law networks. Statistical analysis of network data shows that many networks possess a power-law degree distribution \cite{clauset2009,newman2010,vazquez2002,newman2002b}. Traditionally, this is captured by using the CM and assuming that the probability $p_k$ that a node has $k$ neighbors then scales with $k$ as $p_k\sim C k^{-\tau}$ for some constant $C$ and power-law exponent $\tau>0$.
Many scale-free networks have an exponent $\tau$ between 2 and 3 \cite{albert1999,faloutsos1999,Barabasi2000}, so that the second moment of the degree distribution diverges in the infinite-size network limit.
The exponent $\tau$ is an important characteristic of the network and determines for example the mean distances in the network~\cite{hofstad2005,hofstad2007, newman2001}, or the behavior of various processes on the graph like bond percolation~\cite{callaway2000}, first passage percolation~\cite{bhamidi2010} and an intentional attack~\cite{cohen2001}.
Using the HCM instead of the CM, it no longer suffices to describe the degree distribution $p_k$, but instead assumptions need to made about the joint distribution $p_{k,s}$. In the special case of extremely dense communities, this joint perspective gives rise to the following phenomenon (that we formalize in Section~\ref{sec3}):
\textit{
If the total degree distribution of a network with extremely dense communities follows a power law with exponent $\tau$, the power law of the community sizes has exponent $\gamma=\tau+1$}.
In the household model, where each community is a complete graph, we observe that indeed $\gamma=\tau+1$.
However, real-world network data shows that communities are not always extremely dense, in which case we find that $\gamma\neq \tau+1$.
This is due to the fact that the edge density of communities turns out to decay with community size.

The outline of this paper is as follows. In Section \ref{sec2} we introduce the model and show how generating function techniques can be used to obtain exact large-graph limit results for the giant component.
In Section \ref{sec4} we consider percolation on the HCM and again using generating function techniques we obtain the critical point and the size of a giant percolation cluster. We further investigate the delicate relation between community structure and percolation and show that community structure can both enforce and inhibit percolation.
In Section \ref{sec3} we consider the special case in which the degree distribution follows a power law. It is here that we discover the power-law shift caused by community structure when the communities are \emph{extremely dense}.
In Section \ref{sec5} we apply the HCM to several real-world networks, and we observe two more power-law relations in graphs with communities.
In Section~\ref{sec6} we present conclusions and future research directions.

\section{Model description}\label{sec2}
\begin{figure}[t]
	\centering
	\includegraphics[width=0.13\textwidth]{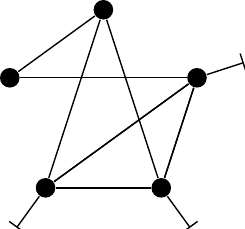}
	\caption{A community with $s=5$ and $k=3$}
	\label{fig:commex}
	\end{figure}

We define the HCM as an extension of the CM, in which each vertex in the CM is replaced by a community \--- some connected graph. We denote community $i$ by $H_i=(F_i,\boldsymbol{d}_i)$. Here $F_i=(V_{H_i},E_{H_i})$ is a graph, defining the shape of the community. The vector $\boldsymbol{d}_i$ counts the number of half-edges attached to each vertex in $F_i$, going out of the community. These half-edges will form the inter-community connections. Let $s_i$ be the size of community $i$, and $k_i$ the number of edges from community $i$ to other communities. Figure~\ref{fig:commex} shows an example of a community with $s=5$ and $k=3$. If we order the vertices clockwise, starting at the vertex in the top of the graph, then $\boldsymbol{d}=(0,1,1,1,0)$.  The number of half-edges attached to vertex $v$ is denoted by $d_v^{\sss{(b)}}$, and referred to as \textit{outside degrees} or inter-community degrees, which define the connections between communities. Similarly, the number of edges from vertex $v$ to other vertices inside the same community is denoted by $d_v^{\sss{(c)}}$, the \textit{inside degrees} or intra-community degrees, the collection of which defines the local connections inside the communities. The degree of vertex $v$ satisfies $d_v=d_v^{\sss{(b)}}+d_v^{\sss{(c)}}$.
As in the CM, the random graph is constructed by picking two half-edges at random, and pairing them. This procedure continues until no half-edges are left. Thus, the edges that connect different communities are formed as in the CM, but the intra-community edges are fixed.

We define the joint distribution $p_{k,s}$ to be the fraction of communities of size $s$ having outside degree $k$. We denote the number of vertices in the random graph by $N$ and the number of communities by $n$. To calculate properties of the random graph, we define several distributions and their probability generating functions (pgfs). Let
	\begin{equation}
	g_p(x,y)=\sum_{k,s}p_{k,s}x^ky^s
	\end{equation}
denote the pgf of the joint size and outside degree distribution of the communities. Introduce the \textit{excess outside degree distribution} by
	\begin{equation}
	q_{k,s}=\frac{(k+1)p_{k+1,s}}{\mean{k}},
	\end{equation}
where $q_{k,s}$ can be interpreted as the probability to arrive in a community of size $s$ and outside degree $k+1$ when traversing a random inter-community edge, including the traversed edge. Here $\mean{k}=\sum_{k,s}kp_{k,s}$ is the expected value of $k$.
Similarly, define
	\begin{equation}
	r_{k,s}=\frac{sp_{k,s}}{\mean{s}}
	\end{equation}
as the probability that a randomly chosen vertex is in a community of size $s$ (including the vertex itself) and has $k$ edges to other communities.
The pgfs for these probability distributions are given by
	\begin{align}
	g_q(x,y)&=\sum_{k,s}q_{k,s}x^ky^s=\frac{1}{\mean{k}}\sum_{k,s}kp_{k,s}x^{k-1}y^s\nonumber \\
	&=\frac{1}{\mean{k}}\frac{\partial g_p(x,y)}{\partial x},\\
	g_r(x,y)&=\sum_{k,s}r_{k,s}x^ky^s=\frac{1}{\mean{s}}\sum_{k,s}sp_{k,s}x^{k}y^s\nonumber \\
	&=\frac{y}{\mean{s}}\frac{\partial g_p(x,y)}{\partial y}.\label{eq:gr}
	\end{align}
We use these pgfs to calculate the size of the largest connected component in the graph.

\subsection{Emergence and size of a giant component}\label{sec:giant}

Let us first explain why the HCM remains amenable for analysis using the generating function technique. Although the HCM is not locally tree-like, the connections between communities are formed as in the CM. Therefore, on the higher level of communities, the HCM is still locally tree-like, and the probability that a half-edge attached to a community forms a self-loop or multiple edge with other communities, tends to zero as $N$ grows large.

Let $u$ be the probability that a community that is reached by traversing a random inter-community edge is not in the giant component, in which case all the communities connected to it cannot be in the giant component either. The $k$ neighboring communities of the reached community are not in the giant component with probability $u^k$. Hence, a community is not in the giant component with probability
	\begin{equation}\label{eq:u}
	u=\sum_{k,s}q_{k,s}u^k=g_q(u,1).
	\end{equation}
Similarly, the probability that a randomly chosen vertex is not in the giant component is $\sum_{k,s}r_{k,s}u^k=g_r(u,1)$. Thus, the size of the largest component $S$ satisfies
	\begin{equation}\label{eq:S}
	S=1-g_r(u,1).
	\end{equation}
A giant component exists if and only if $S>0$. We can see that $S=0$ if and only if $u=1$, so $S>0$ for $u<1$. Furthermore,~\eqref{eq:u} shows that $u$ is the extinction probability of a branching process with degree distribution $\sum_sq_{k,s}$  and expected offspring
	\begin{equation}\label{eq:condgiant}
	\sum_{k,s}kq_{k,s}=\frac{\mean{k(k-1)}}{\mean{k}}.
	\end{equation}
It is well known that the condition $u<1$ for the existence of a giant component is equivalent to the expected offspring being larger than one, or $\mean{k^2}-2\mean{k}>0$~\cite{molloy1995}.
This is the same condition as for the CM with offspring $(p_k)_{k\geq 0}$, so that the point at which the giant component emerges is only determined by the inter-community connections. Here we have made the implicit assumption that $\mean{s}<\infty$, to be able to use~\eqref{eq:gr}. As long as $\mean{s}<\infty$, the community sizes have no influence on the point at which the giant component emerges.

In general, $k$ and $s$ can be dependent. If $k$ and $s$ were independent, that is $p_{k,s}=p_kp_s$ and
	\begin{equation}
	u={\mean{k}}^{-1}\sum_{k,s}kp_{k}p_su^{k-1}={\mean{k}}^{-1}\sum_k kp_k u^{k-1},
	\end{equation}
then
 	\begin{equation}
 	S=1-{\mean{s}}^{-1}\sum_{k,s}sp_{k}p_{s}u^k=1-\sum_k p_ku^k.
 	\end{equation}
These equations are the same as for the CM with degree distribution $(p_k)_{k\geq 0}$.
Therefore, when $k$ and $s$ are independent, also the size of the giant component is entirely defined by the inter-community connections.
However, in real-world networks, $k$ and $s$ are likely to be dependent. Independence would imply that a small community has the same probability of having a large number of edges towards other communities as a large community. In most real-world examples it is more likely that every vertex inside a community has some edges towards other communities, so a larger community has a larger probability of $k$ being large than a smaller community, in which case $k$ and $s$ are positively correlated.

It is also possible to calculate the sizes of the other components, when there is no giant component. Let $h_q(z)$ be the pgf of the number of vertices accessible from a randomly chosen inter-community edge. When the edge reaches a community of size $s$, this adds $s$ vertices to the component, which contributes a factor $z^s$. Furthermore, if the community has $k$ other outgoing edges, then each of these edges will generate a component with pgf $h_q(z)$. This gives
	\begin{equation}\label{eq:hq}
	h_q(z)=\sum_{k,s}q_{k,s}z^sh_q(z)^k=g_q(h_q(z),z).
	\end{equation}
Now we derive $h_p(z)$, the pgf of the size of the component of a uniformly chosen vertex.
When a uniformly chosen vertex is in a community of size $s$ and outside degree $k$, the members of the community add $z^s$ to the pgf. Each half-edge generates a component with pgf $h_q(z)$. This gives
	\begin{equation}\label{eq:hp}
	h_p(z)=\sum_{k,s}r_{k,s}z^sh_q(z)^k=g_r(h_q(z),z).
	\end{equation}
The mean component size is given by $h'_p(1)$. Differentiating~\eqref{eq:hq} and~\eqref{eq:hp} yields
	\begin{align}
	h'_q(1)&=\frac{1}{\mean{k}}\frac{\partial g_p}{\partial x^2}(1,1)h'_q(1)+\frac{1}{\mean{k}}\frac{\partial g_q}{\partial xy}(1,1)\nonumber\\
	&=\frac{\mean{k(k-1)}}{\mean{k}}h'_q(1)+\frac{\mean{ks}}{\mean{s}},\\
	h'_p(1)&=\frac{\partial g_r}{\partial x}(1,1)h'_q(1)+\frac{\partial g_r}{\partial y}(1,1)\nonumber \\
	&=\frac{\mean{ks}}{\mean{s}}h'_q(1)+\frac{\mean{s^2}}{\mean{s}}.\label{eq:hp1der}
	\end{align}
These equations together define $h'_q(1)$ and $h'_q(1)$ and some
rewriting yields
	\begin{align}
	\label{eq:hq1}
	h'_q(1)&=\frac{\mean{ks}}{2\mean{k}-\mean{k^2}},\\
	\label{eq:hp1}
	h'_p(1)&=\frac{\mean{s^2}}{\mean{s}}+\frac{\mean{ks}^2}{\mean{s}(2\mean{k}-\mean{k^2})}.
	\end{align}
when $2\mean{k}-\mean{k^2}>0$.
The first term in~\eqref{eq:hp1} is the expected size of the component in which the randomly chosen vertex lies. The second term equals the expected size of the components attached to the community of the randomly chosen vertex.
The expected component size is infinite if the giant component emerges (if $2\mean{k}-\mean{k^2}>0$). Equation~\eqref{eq:hp1} shows that $\mean{ks}=\infty$ or $\mean{s^2}=\infty$ also give an infinite expected component size. However, the condition for the giant component to emerge does not involve $s$. Thus, it is possible to have an infinite expected cluster size without having a giant component. Then the expected cluster size is infinite, but still small compared to the total  number of vertices in the graph.
One example of a random graph with no giant component but an infinite expected cluster size is a graph with $k=0$ for all communities (all communities are isolated), and $p_s=Cs^{-\alpha}$ for some $\alpha\in(2,3)$. This example has $\mean{s^2}=\infty$, hence by~\eqref{eq:hp1} the expected cluster size of a randomly chosen vertex is infinite. Since $k=0$ for all communities, clearly condition~\eqref{eq:condgiant} is not met, and there is no giant component. The size of the largest component in this example is the size of the largest community $s_{\text{max}}$. If there are $n$ communities, then $s_{\text{max}}\sim n^{1/(\alpha-1)}$~\cite{newman2003book}. The total number of vertices in the HCM with $n$ communities goes to $n\mean{s}$. Thus, the fraction of vertices in the largest component behaves like
	\begin{equation}
	\frac{n^{1/(\alpha-1)}}{n\mean{s}}=\frac{1}{\mean{s}}n^{\frac{2-\alpha}{\alpha-1}}\to 0,
	\end{equation}
as $n\to\infty$. The same phenomenon occurs when the community size distribution is the same, but we add some edges between communities. So as long as $\mean{k^2}-2\mean{k}<0$, there is no giant component, even though the expected cluster size is infinite.

\section{Percolation}\label{sec4}
In this section we consider bond percolation on the HCM, where each edge is removed independently with probability $1-\phi$. We are interested in the size of the giant component after removing the edges. We calculate this in a similar way as we computed the size of the giant component before percolation. Next to the joint distribution $p_{k,s}$ we define the distribution $p_H$ that denotes the fraction of communities of type $H$.

Deleting each edge with probability $1-\phi$ is the same as first deleting only the intra-community edges with probability $1-\phi$ and then the inter-community edges also with probability $1-\phi$. Thus, first delete each edge inside the communities with probability $1-\phi$. After this procedure, some communities may have split into several connected components. However, these components still form connections as in the CM. Hence, after percolation inside the communities, we again have an HCM. The communities in the new HCM are the connected components of the percolated communities.

When entering a percolated community, the percolated community no longer needs to be connected, so that the $k$ edges to other communities are not always reached. To account for this, we introduce $f(H,v,l,\phi)$, the probability that after percolation, the connected component of community $H$ containing vertex $v$ still has $l$ outgoing edges. Let $t_k^{\sss{(\phi)}}$ be the probability that a randomly chosen vertex is in a percolated community with $k$ edges to other communities. Vertex $v$ in community $H$ is chosen with probability $p_H/{s_H}$. The probability that $v$ is connected to $k$ half-edges is given by $f(H,v,k,\phi)$.
Hence, $t_k^{\sss{(\phi)}}$ is given by
	\begin{equation}
	t_k^{\sss{(\phi)}}=\sum_{H}\sum_{v\in V_H}\frac{1}{s_H}p_Hf(H,v,k,\phi).
	\end{equation}
Let $q_k^{\sss{(\phi)}}$ denote the probability that a percolated community reached by traversing a randomly chosen inter-community edge has $k$ edges towards other communities. The probability of arriving in vertex $v$ of community $H$ is proportional to $p_Hd_v^{\sss{(b)}}$. Then the probability that $v$ is inside a percolated community with $k$ other outgoing edges is $f(H,v,k+1,\phi)$, so that
	\begin{equation}
	q_k^{\sss{(\phi)}}\propto\sum_{H}\sum_{v\in V_H}d_v^{\sss{(b)}}p_Hf(H,v,k+1,\phi).
	\end{equation}
This gives
	\begin{equation}
	\begin{aligned}[b]
	q_k^{\sss{(\phi)}}&=\frac{\sum_{H}\sum_{v\in V_H}d_v^{\sss{(b)}}p_Hf(H,v,k+1,\phi)}{\sum_l\sum_{H}\sum_{v\in V_H}d_v^{\sss{(b)}}p_Hf(H,v,l,\phi)}\\
	&=\frac{\sum_{H}\sum_{v\in V_H}d_v^{\sss{(b)}}p_Hf(H,v,k+1,\phi)}{\mean{k}}.
	\end{aligned}
	\end{equation}
Define $g_{q^{\sss{(\phi)}}}(z)$ and $g_{t^{\sss{(\phi)}}}(z)$ as the pgf of $q_k^{\sss{(\phi)}}$ and $t_k^{\sss{(\phi)}}$ respectively, which will be used to calculate the size of the giant percolation cluster.

\subsection{Giant percolation cluster}

After percolating the edges inside communities, we percolate the edges between communities. Since the inter-community edges are paired as in the CM, percolation on these edges is similar to percolation on the CM~\cite{janson2008,callaway2000}. We remove each half-edge of a community with probability $1-\sqrt{\phi}$. Then an edge is removed when at least one of the two half-edges that form the edge is removed. Thus, an edge is removed with probability $2(1-\sqrt{\phi})\sqrt{\phi}+(1-\sqrt{\phi})^2=1-\phi$, as required.

Given that the number of half-edges that are attached to a community before percolating is $k$, the number of half-edges after percolating has a binomial distribution with parameters $(k,\sqrt{\phi})$. If we denote by $X^{\sss{(\phi)}}$ and $Q^{\sss{(\phi)}}$ the number of half-edges of a community entered via a randomly chosen edge after and before percolation, respectively, then
	\begin{align}
	g_{X^{\sss{(\phi)}}}(z)&=\sum_{l=1}^\infty z^l\mathbb{P}(X^{\sss{(\phi)}}=l)\nonumber\\
	&=\sum_{l=1}^\infty\sum_{k=1}^\infty z^l\mathbb{P}(X^{\sss{(\phi)}}=l\mid Q^{\sss{(\phi)}}=k)q^{\sss{(\phi)}}_k\nonumber\\
	&=\sum_{k=1}^\infty q^{\sss{(\phi)}}_k\sum_{l=1}^\infty z^l\mathbb{P}(\text{Bin}(k,\sqrt{\phi})=l)\nonumber\\
	&=\sum_{k=1}^\infty q^{\sss{(\phi)}}_k(1-\sqrt{\phi}+\sqrt{\phi}z)^k\nonumber\\
	&=g_{q^{\sss{(\phi)}}}(1-\sqrt{\phi}+\sqrt{\phi}z).
	\end{align}
Here $\mathbb{P}(\text{Bin}(n,p)=i)={n\choose i} p^i(1-p)^{n-i}$ is the probability that a binomial random variable with parameters $(n,p)$ takes the value $i$.

As in the previous section, let $u^{\sss{(\phi)}}$ denote the probability that a vertex reached from a uniformly chosen half-edge is not in the giant component after percolation. One possibility is that the half-edge that we follow links to a deleted half-edge, which happens with probability $1-\sqrt{\phi}$. In this case, the half-edge does not lead to the giant component with probability one. If the chosen half-edge links to a half-edge that was not deleted (which happens with probability $\sqrt{\phi}$), then it leads to a vertex inside a percolated community. This community is not in the giant component if all of its half-edges do not link to the giant component, which happens with probability $g_{X^{\sss{(\phi)}}}(u^{\sss{(\phi)}})$. This results in
	\begin{equation}\label{eq:uphi}
	u^{\sss{(\phi)}}=\big(1-\sqrt{\phi}\big)+\sqrt{\phi}g_{q^{\sss{(\phi)}}}\left(1-\sqrt{\phi}+\sqrt{\phi}u^{\sss{(\phi)}}\right).
	\end{equation}
A randomly chosen vertex is not in the giant component when all the half-edges of its percolated community do not link to the giant component. Hence, the size of the giant component after percolation $S^{\sss{(\phi)}}$ satisfies
	\begin{equation}\label{eq:Sphi}
	1-S^{\sss{(\phi)}}=g_{t^{\sss{(\phi)}}}(1-\sqrt{\phi}+\sqrt{\phi}u^{\sss{(\phi)}}).
	\end{equation}
Solving equations~\eqref{eq:uphi} and~\eqref{eq:Sphi} together gives the size of the giant component after percolation.

\subsection{Percolation transition point}\label{sec:pic}

To find the percolation transition point, we view the number of communities that can be reached by traversing a random inter-community edge (excluding the traversed edge) as a branching process. The offspring distribution of this branching process is the distribution of the number of half-edges attached to a percolated community reached by traversing a random edge. The expected number of such half-edges after percolation inside the communities is $\mean{q^{\sss{(\phi)}}}$. When we then delete the inter-community edges with probability $1-\phi$, the mean number of half-edges of a community reached by traversing a random edge (excluding the traversed edge) is $\phi\mean{q^{\sss{(\phi)}}}$.
Hence, we view the number of communities that can be reached from a random edge as a branching process with mean offspring $\phi\mean{q^{\sss{(\phi)}}}$. This immediately shows that the percolation transition point $\phi_c$ is when $\phi_c\mean{q^{\sss{(\phi_c)}}}=1$, so that
	\begin{equation}\label{eq:pic}
	\phi_c=
	\frac{\mean{k}}{\sum_{H}\sum_{v\in V_H}\sum_k kd_v^{\sss{(b)}}p_Hf(H,v,k+1,\phi_c)}.
	\end{equation}

Equations~\eqref{eq:Sphi} and~\eqref{eq:pic} show that the critical percolation value as well as the size of the giant component after percolation depend on the shapes of the communities. Appendix~\ref{sec:exampleperc} shows that introducing a community structure may either increase or decrease the critical percolation value as well as the size of the giant component after percolation, and can even lead to non-convex percolation curves.

An interesting example is the case of highly connected communities. These communities are robust against percolation compared to the inter-community connections. Then, close to the critical value $\phi_c$, these communities will still be connected after percolation, and $f(H,v,l,\phi)\approx1$ for $l=k$. Then~\eqref{eq:pic} reduces to $\phi_c=\mean{k}/(\mean{k^2}-\mean{k})$, as in the standard CM. Thus, if communities are highly connected compared to the inter-community edges, the critical percolation value is entirely determined by the inter-community edges.

\section{Power-law community sizes}\label{sec3}

A potentially crucial property observed in many networks
is that the community size distribution appears to
have a power-law form over some significant range~\cite{boguna2004, clauset2004, guimera2003, radicchi2004}.
We now assume that both the degree and the community  size distributions obey power laws with exponents $\tau$ and $\gamma$, respectively. Typical values reported in the literature are $2 \leq \tau\leq 3$ and $1\leq \gamma\leq 3$~\cite{fortunato2010}.

\begin{figure}[t]
	\centering
	\includegraphics[width=0.15\textwidth]{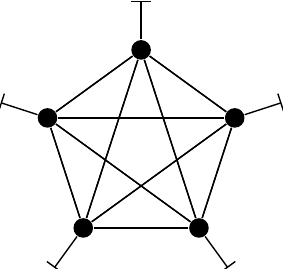}
	\caption{A household community with $s=5$}
	\label{fig:household}
	\end{figure}

\paragraph{Household communities.}
An extreme community structure is that of household communities~\cite{trapman2007}, in which all communities are complete graphs.
Each vertex inside the community has outside degree one. Figure~\ref{fig:household} shows an example of a household community with $s=5$.

\begin{figure}[t]
	\centering
	\includegraphics[width=0.4\textwidth]{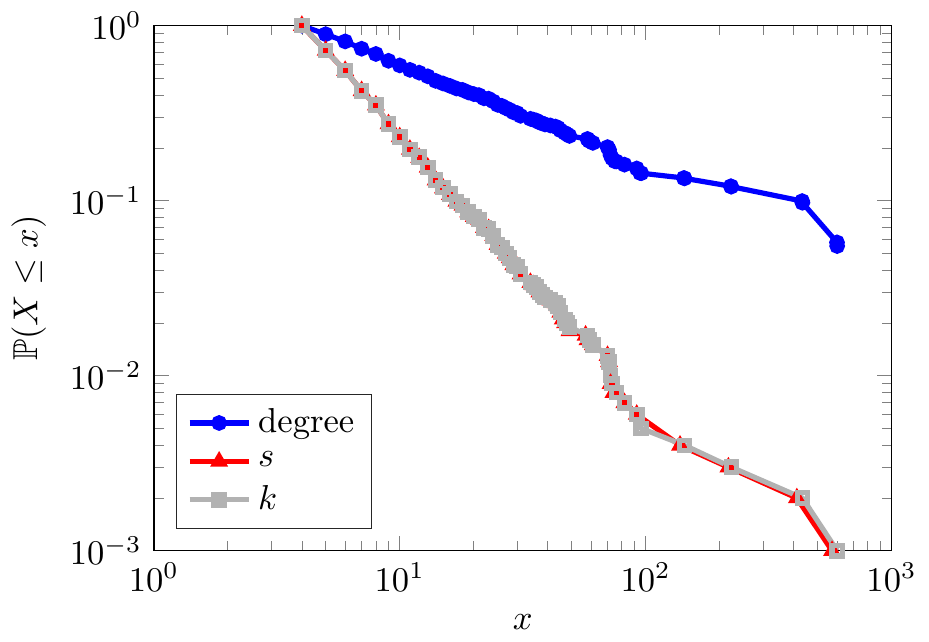}
	\caption{The degree distribution of a household model follows a power law with a smaller exponent than the community size distribution and outside degree distribution}
	\label{fig:householdpl}
	\end{figure}

In a household community, $k=s$, hence $p_{k,s}=0$ if $k\neq s$.
Suppose the distribution of the community sizes follows a power law with exponent $\gamma$, $p_{k,k}=Ck^{-\gamma}$. Then the outside degrees also follow a power-law distribution with exponent $\gamma$. Now we derive $(\hat{p}_k)_{k\geq 0}$, the degree distribution of the HCM with this household structure.
For a vertex in the household model to have degree $k$, it must be in a community of size $k$. Furthermore, there are $k$ of such vertices inside each community, so that
	\begin{equation}
	\hat{p}_k=\frac{kp_{k,k}}{\sum_{i=1}^\infty ip_{i,i}}=\frac{kC_k^{-\gamma}}{\mean{s}}=C_2k^{-\gamma+1}.
	\end{equation}
Thus, the degree distribution of the graph with household communities again obeys a power law but with exponent $\tau=\gamma-1$, as observed in~\cite{trapman2007}.
We call this phenomenon a \textit{power-law shift}, because the edges out of a community have a smaller degree distribution than the individual edges
(see Figure~\ref{fig:householdpl}).

\paragraph{Extremely dense communities.}
In real-world networks, communities may not be complete, but a power-law shift also occurs in networks with an incomplete but extremely dense community structure.
In an extremely dense community, many edges are contained in communities. Let $e_{\text{in}}$ denote the number of edges inside a community. We assume that there exists $\varepsilon>0$ independent of the number of communities $n$  such that for each community $H$,
	\begin{equation}\label{eq:dense}
	e_{\text{in}}\geq\varepsilon s(s-1).
	\end{equation}
In this case, every community of size $s$ contains a positive fraction of the edges that are present in a complete graph of the same size. Note that the household model gives $\varepsilon=\frac12$.

Since the power-law shift states that the outside degrees of the communities are `small', we need the outside degree of individual vertices to be small as well. Thus, we assume that there exists a $K<\infty$ such that for all vertices
	\begin{equation}\label{eq:K}
	d_v^{\sss{(b)}}\leq Ks_i.
	\end{equation}
Note that this implies that $k\leq Ks^2$ for every community. 
Using assumptions~\eqref{eq:dense} and~\eqref{eq:K} we show that a power-law shift occurs.

Suppose that the community size distribution follows a power law with exponent $\gamma$. Denote the cumulative degree distribution by $P_i=\sum_{j\leq i}\hat{p}_j$.
Since the maximal inside degree of a vertex is $s-1$, and by~\eqref{eq:dense} the average inside degree of a vertex is greater than or equal to $\varepsilon(s-1)$, at least a fraction of $\varepsilon$ vertices in any community have inside degree at least $\varepsilon (s-1)$
Thus, a vertex inside a community of size $i/\varepsilon+1$ has probability of at least $\varepsilon$ to have inside degree at least $\varepsilon(i/\varepsilon+1-1)=i$. Hence, $1-P_i$ is bounded from below by $\varepsilon$ times the probability of choosing a vertex in a community of size at least $i/\varepsilon+1$.
The probability that a randomly chosen vertex is in a community of size $j$ is given by $\sum_kr_{k,j}$. This yields
	\begin{align}\label{eq:pllow}
	1-P_j&\geq  \sum_{i\geq j}\sum_k r_{k,i/\varepsilon+1}\varepsilon\nonumber\\
	&= \sum_{i\geq j}\left(\frac i\varepsilon+1\right) \sum_k p_{k,i/\varepsilon+1}\frac{1}{\mean{s}}\varepsilon\nonumber\\
	&\approx Cj^{-\gamma+2}.
	\end{align}
Furthermore, given the distribution of the community sizes, $1-P_j$ is maximal when all communities are complete graphs, and every vertex has $Ks$ half-edges attached to it. Then each vertex in a community of size $s$ has degree $s-1+Ks$.  Hence, to choose a vertex with degree at least $j$, we have to choose a vertex inside a community of size at least $(j+1)/(K+1)$. Then
	\begin{align}\label{eq:plup}
	1&-P_j\leq \sum_{i\geq \frac{j+1}{K+1}} \sum_k r_{k,i}\nonumber\\
	&= \sum_{i\geq \frac{j+1}{K+1}}i\sum_k p_{k,i}\frac{1}{\mean{s}}
	=\frac{c}{\mean{s}}\left(\frac{j+1}{K+1}\right)^{-\gamma+2}.
	\end{align}
Combining~\eqref{eq:pllow} and~\eqref{eq:plup} shows that the degree distribution follows a power law with exponent $\tau=\gamma-1$.
In other words, when the community size distribution of a network with extremely dense communities follows a power law with exponent $\gamma$, the power law of the degrees has exponent $\tau=\gamma-1$.

Under a more strict assumption on the inter-community degrees
	\begin{equation}\label{eq:Kstrict}
	d_v^{\sss{(b)}}\leq K,
	\end{equation}
we can also relate the power-law exponent of the intra-community degrees to the exponent of the degree distribution. Assumption~\eqref{eq:Kstrict} implies $k\leq sK$, and therefore if the community size distribution follows a power law with exponent $\gamma=\tau+1$, then the distribution of the community outside degrees cannot have a power-law distribution with exponent smaller than $\gamma$.
Suppose we want to construct a graph where the degree distribution follows a power law with exponent $\tau\in(2,3)$.
One possibility to construct such a graph is to use the CM.
However, the CM with $\tau\in(2,3)$ has probability zero to create a simple graph.
Another way to construct a graph with this degree distribution is to use the HCM with extremely dense communities of power-law size with exponent $\tau+1$. The outside degrees of the communities then follow a power law with exponent at least $\tau+1\geq 3$. Since the outside degrees are paired according to the CM, the probability that the resulting graph is simple, will be larger than zero in the limit of infinite graph size. Thus, the HCM is able to construct a simple graph with exponent $\tau\in(2,3)$.

Another interesting application of this power-law shift is in the critical percolation value. It is well known that the critical percolation value $\phi_c=0$  for a CM with $\tau\in(2,3)$~\cite{molloy1995}. Section~\ref{sec:pic} showed that for highly connected communities, the critical percolation value is entirely defined by the inter-community degrees. Since the inter-community degrees have exponent larger than $3$, the HCM is able to construct random graphs with $\tau\in(2,3)$ and $\phi_c>0$. This shows that the HCM with extremely dense communities is in another universality class than the CM.

\begin{figure}[tb]
	\centering
	\subfloat[]{
		\centering
		\includegraphics[width=0.15\textwidth]{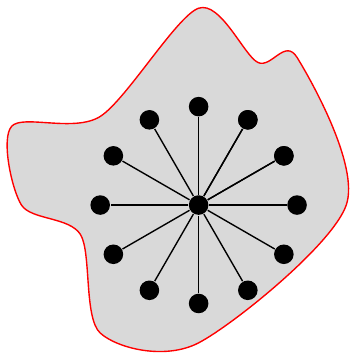}
		\label{fig:hub1}
	}
	\subfloat[]{
		\centering
		\includegraphics[width=0.15\textwidth]{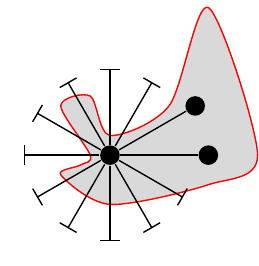}
		\label{fig:hub2}
	}
	\caption{(a): a hub with all its neighbors completely inside a community (shaded area).(b): A hub with only a few neighbors inside the same community}
	\label{fig:hub}
\end{figure}

\paragraph{The role of hubs.}
A power-law degree distribution implies the existence of hubs: nodes with a very high degree. We now show that this can conflict with assumptions~\eqref{eq:dense} and~\eqref{eq:K}.
Since every vertex is inside a community in the HCM, the hubs also need to be assigned to some community. In these communities, hubs can have several roles, as observed in~\cite{guimera2005}. There are two possibilities, as shown in Figure~\ref{fig:hub}.  When most neighbors of the hub are also inside the community as in Figure~\ref{fig:hub1}, then the hub is in a very large community. Assumption~\eqref{eq:dense} states that most neighbors of the hub should also be connected to one another, and thus also have a high degree.  However, in real-world networks this might not be realistic. For example, when one person in a social network has many friends, this does not mean that most of these friends are friends with one another. Hence, putting most neighbors of a hub inside the same community can create communities that are not dense. The other possibility (see Figure~\ref{fig:hub1}) is to have only a small fraction of the neighbors inside a community. However, now the outside degree of the hub is large, which may contradict assumption~\eqref{eq:K} when the hub is in a small community. Therefore, the existence of hubs conflicts with the assumption of extremely dense communities.

\section{Real-world networks}\label{sec5}
In this section, we apply the HCM to four different data sets: an \textsc{Amazon} co-purchasing network~\cite{yang2015}, the \textsc{Gowalla} social network~\cite{cho2011}, a network of relations between English words~\cite{miller1998} and a \textsc{Google} web graph~\cite{leskovec2009}. To extract the community structure of the networks, we use the Infomap community detection method~\cite{Rosvall2008}, a community detection method that performs well on several benchmarks~\cite{lancichinetti2009}. Table~\ref{tab:alphabeta} shows that equation~\eqref{eq:S} identifies the size of the giant component almost perfectly, in contrast to the value calculated by the CM.
	\begin{table}[tb]
  	\centering
  	\caption{Several characteristics of four different data sets}
  	\begin{ruledtabular}
    	\begin{tabular}{lrrrr}
 	\textbf{} & \textbf{\textsc{Amazon}} &{\textbf{\textsc{Gowalla}}} &{\textbf{WordNet}} & \textbf{\textsc{Google}} \\
   	\hline
              	$S$ (data) & 1,000 & 1,000 &     0,994  & 0,977 \\
              	$S$ (HCM) & 1,000 & 1,000 & 0,994 & 0,978 \\
    		$S$ (CM) & 0,999 & 0,993 & 0,999 & 0,997 \\
      		$\gamma$ &3,84  & 2,44  & 3,23  & 2,58 \\
    		$\tau$   &3,59  & 2,48  & 2,82  & 2,73 \\
        		$\alpha$ & 0,15  & 0,31  & 0,21  & 0,21 \\
    		$\beta$& 1,14  & 1,18  & 1,28  & 1,24 \\
      		$\gamma/\alpha-1 $  &25,04 & 6,85  & 14,17 & 11,20 \\
    	\end{tabular}%
    	\end{ruledtabular}
  	\label{tab:alphabeta}%
	\end{table}%

\begin{figure*}[t]
	\centering
	\subfloat[]{
		\centering
		\includegraphics[width=0.35\textwidth]{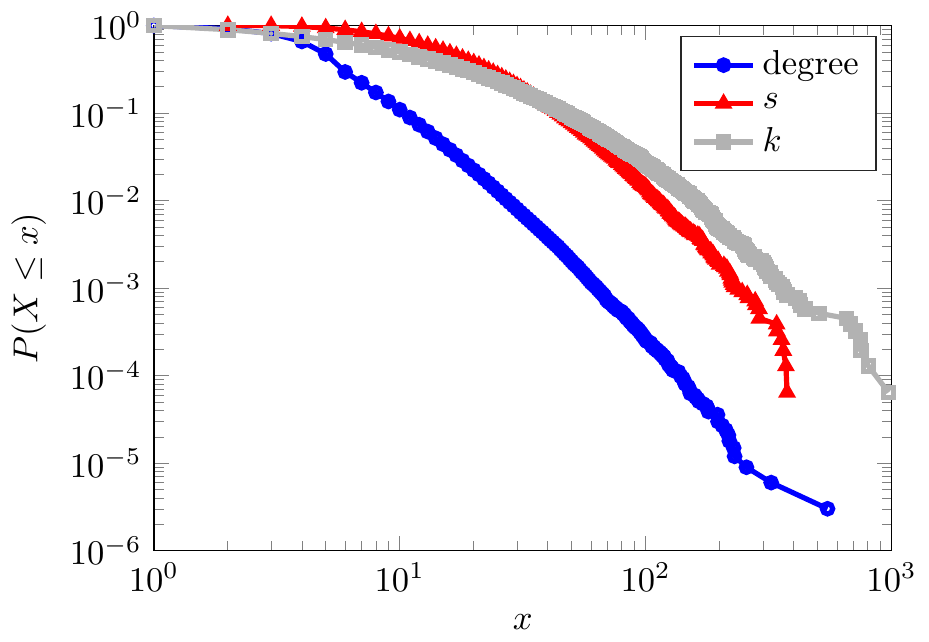}
		
		\label{fig:amazonpl}
	}
	\hspace{0.2cm}
	\subfloat[]{
		\centering
		\includegraphics[width=0.35\textwidth]{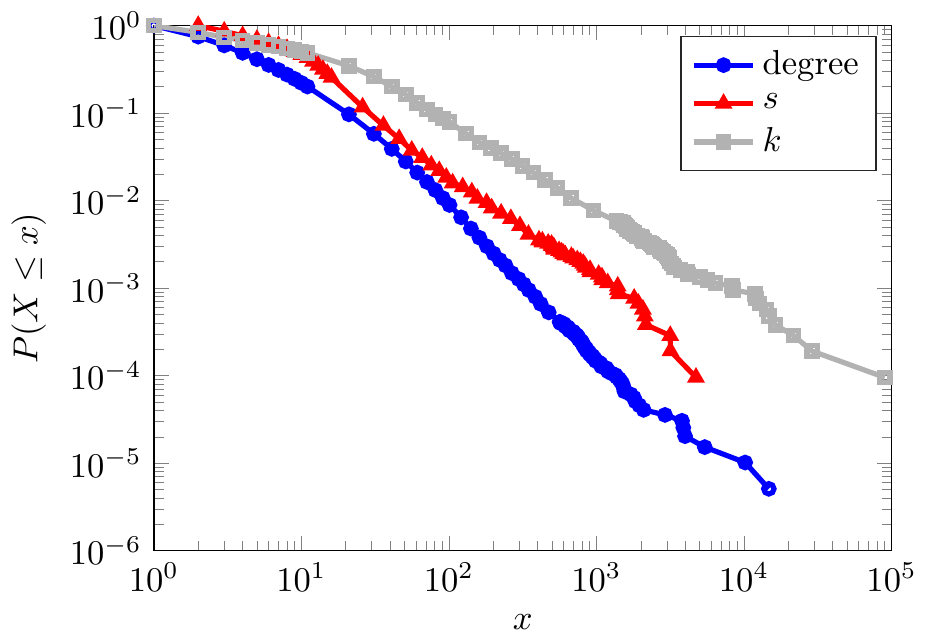}
		
		\label{fig:dlbppl}
	}
	
	\subfloat[]{
		\centering
		\includegraphics[width=0.35\textwidth]{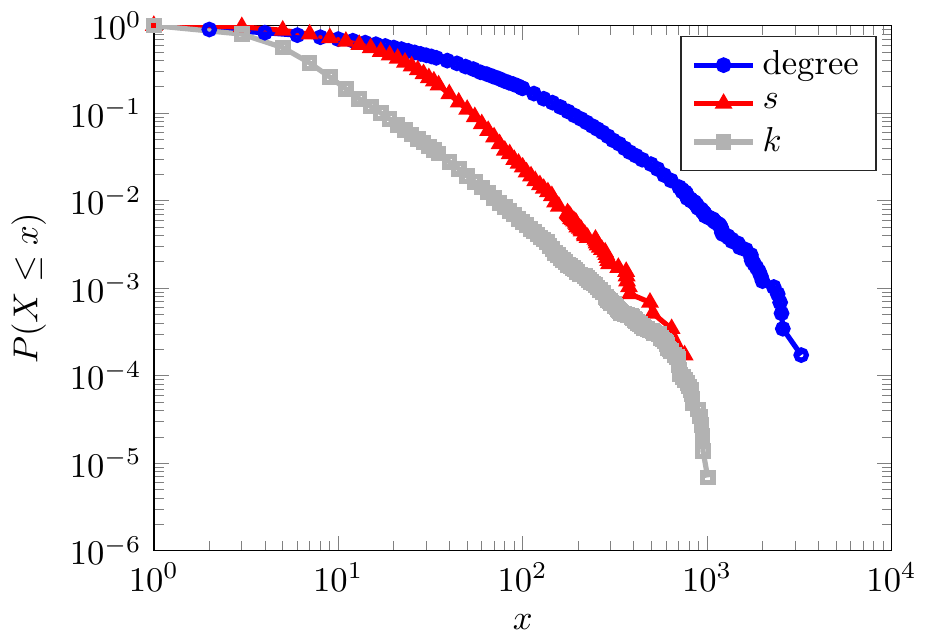}
		
		\label{fig:condpl}
	}
	\hspace{0.2cm}
		\subfloat[]{
		\centering
		\includegraphics[width=0.35\textwidth]{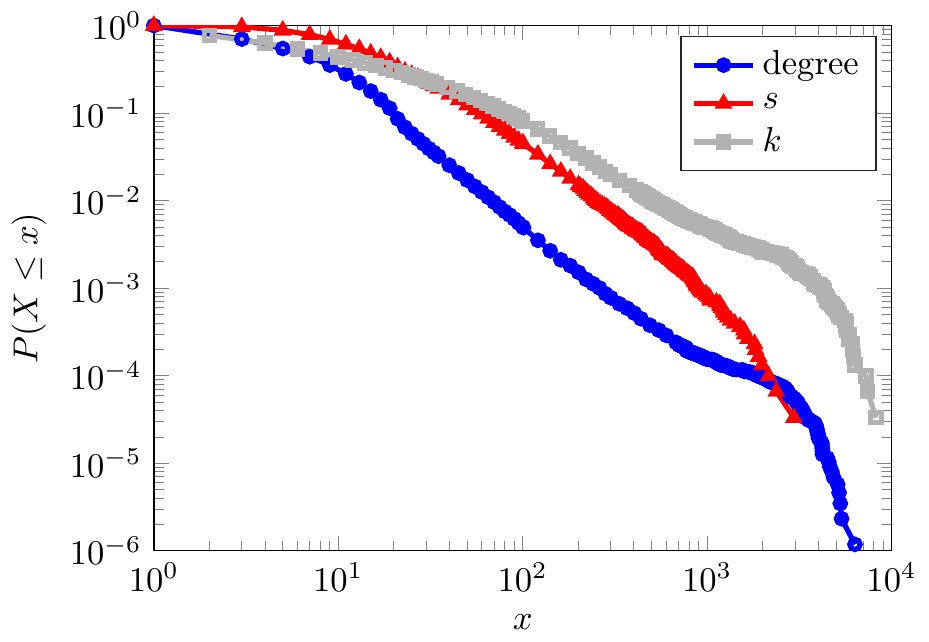}
		
		\label{fig:googlepl}
	}
	\caption{Power-law relations in real-world networks. a) \textsc{Amazon} co-purchasing network, b) \textsc{Gowalla} social network, c) English word relations, d) \textsc{Google} web graph. Both the degrees and the community sizes $s$ follow a power law, as well as the inter-community degrees $k$. }
	\label{fig:plshift}	
\end{figure*}
\begin{figure*}[t]
	\subfloat[]{
		\centering
		\includegraphics[width=0.35\textwidth]{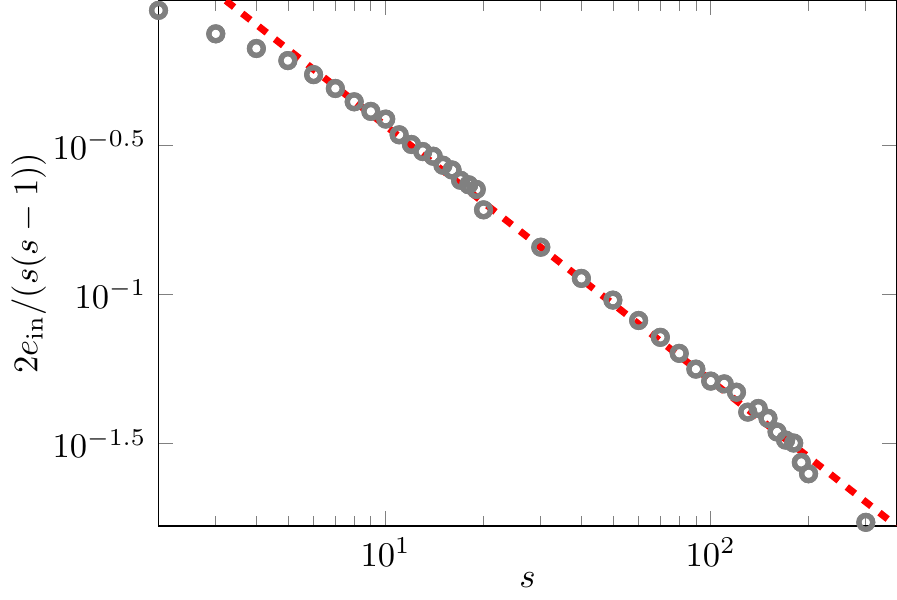}
		
		\label{fig:amade}
	}
	\hspace{0.2cm}
	\subfloat[]{
		\centering
		\includegraphics[width=0.35\textwidth]{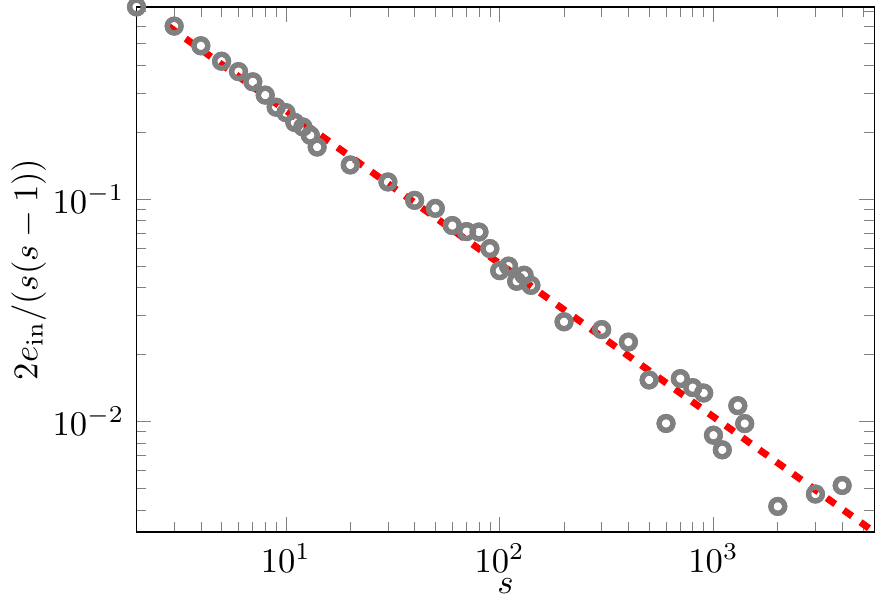}
		
		\label{fig:astrode}
	}
	
	\subfloat[]{
		\centering
		\includegraphics[width=0.35\textwidth]{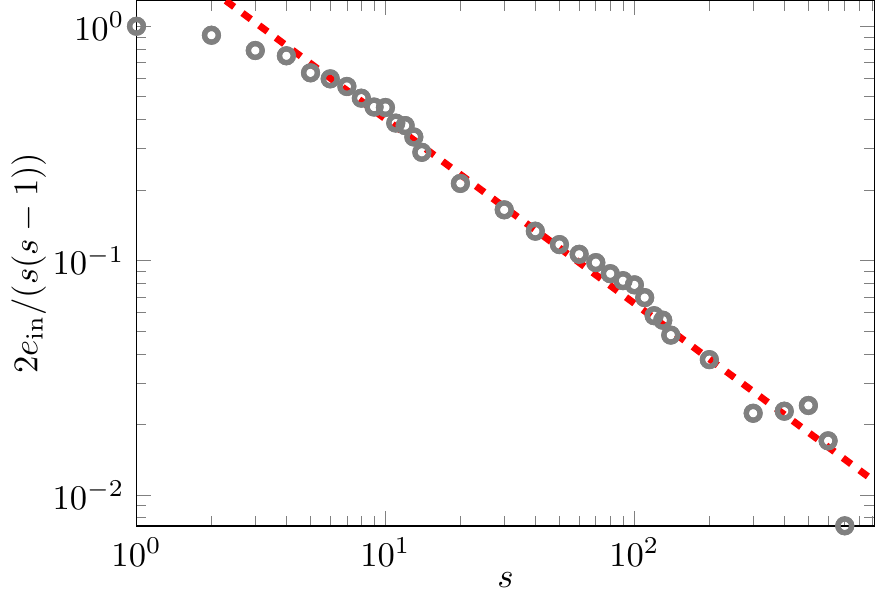}
		
		\label{fig:condde}
	}
	\hspace{0.2cm}
		\subfloat[]{
		\centering
		\includegraphics[width=0.35\textwidth]{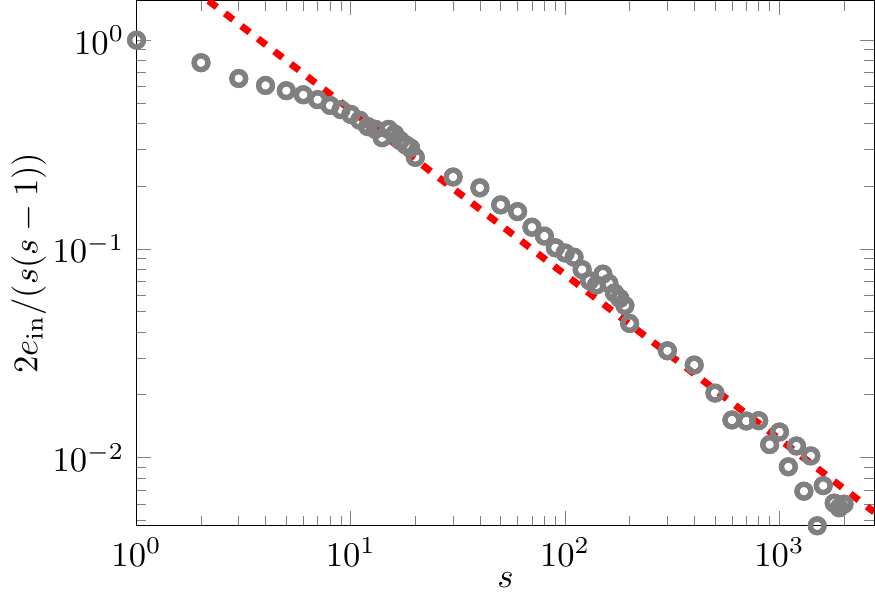}
		
		\label{fig:googlede}
	}
	\caption{The denseness of a community $2e_{\text{in}}/(s^2-s)$ has a power-law relation with the community size $s$. a) \textsc{Amazon} co-purchasing network, b) \textsc{Gowalla} social network, c) English word relations, d) \textsc{Google} web graph.}
	\label{fig:pldense}	
\end{figure*}

Figure~\ref{fig:plshift} illustrates the power laws for these data sets.
Table~\ref{tab:alphabeta} presents values of the power-law exponents $\tau$ and $\gamma$, estimated by the method of Clauset et al.~\cite{clauset2009}.
We see that a power-law shift is less pronounced than in the stylized household model, if existing at all.
This indicates that the communities in the data sets do not have the intuitive dense structure. Thus, we test assumption~\eqref{eq:dense}.
The maximum number of edges inside a community is obtained if the community is a complete graph, in which case $e_{\text{in}}=\frac12 s(s-1)$. Dividing~\eqref{eq:dense} by $s(s-1)/2$ gives $\frac{e_{\text{in}}}{s(s-1)/2}\geq2\varepsilon$. This fraction measures how dense a community is.
Figure~\ref{fig:pldense} plots $s$ against the average value of $2e_{\text{in}}/(s^2-s)$. For all networks, this fraction is not independent of $s$. Larger communities are less dense than smaller communities. Therefore, the large communities do not satisfy the intuitive picture of an extremely densely connected subset, even though the density within communities is much higher than that in the entire network. This is a similar observation as in~\cite{leskovec2009}, where the authors discover that most real-world networks have a strongly connected core, which consists of several interconnected communities that are hard to distinguish. The core is connected to the periphery, some isolated, densely connected small communities. This structure could explain the dependence of the density of the communities on $s$. The large communities that are not very dense, are part of the core, whereas the small communities are the more isolated parts of the network.
Another interesting property of the community structures in Figure~\ref{fig:pldense} is the power-law relation between the community sizes and their densities, $e_{\text{in}}\approx cs^{\alpha+1}$. In assumption~\eqref{eq:dense}, we assume that $\alpha=1$. However, Table~\ref{tab:alphabeta} shows that the example data sets have $\alpha<1$. For this reason, we replace~\eqref{eq:dense} by
	\begin{equation}
	e_{\text{in}}\geq\varepsilon s(s-1)^\alpha.
	\end{equation}
Now~\eqref{eq:plup} still holds, but~\eqref{eq:pllow} needs to be modified. The average inside degree of a vertex now is $\varepsilon(s-1)^\alpha$. Since the maximum inside degree is $s-1$, there are at least $\varepsilon(s-1)^{\alpha-1}$ vertices of degree at least $\varepsilon(s-1)^\alpha$. A similar analysis as~\eqref{eq:pllow} yields
	 \begin{align}\label{eq:Pjnew}
 	1-P_j&\geq  \sum_{i\geq j}\sum_k r_{k,(i/\varepsilon)^{(1/\alpha)}+1}\varepsilon\left(i/\varepsilon\right)^{(\alpha-1)/\alpha}\nonumber\\
 	&=\frac{\varepsilon}{\mean{s}} \sum_{i\geq j}\Big(\Big(\frac i\varepsilon\Big)^{(1/\alpha)}+1\Big)^{-\gamma+1}\left(i/\varepsilon\right)^{(\alpha-1)/\alpha}\nonumber\\
 	&\approx Cj^{-\gamma/\alpha+2}.
 	\end{align}
Together with~\eqref{eq:plup}, this shows that the exponent $\tau$ of the degree distribution satisfies $\tau\in[\gamma-1,\frac{\gamma}{\alpha}-1]$. Table~\ref{tab:alphabeta} shows several values of $\tau$, $\gamma$ and $\gamma/\alpha-1$. We see that indeed $\tau\in[\gamma-1,\frac{\gamma}{\alpha}-1]$ in the example data sets. However, the interval may be quite wide.

\begin{figure*}[t]
	\subfloat[]{
		\centering
		\includegraphics[width=0.35\textwidth]{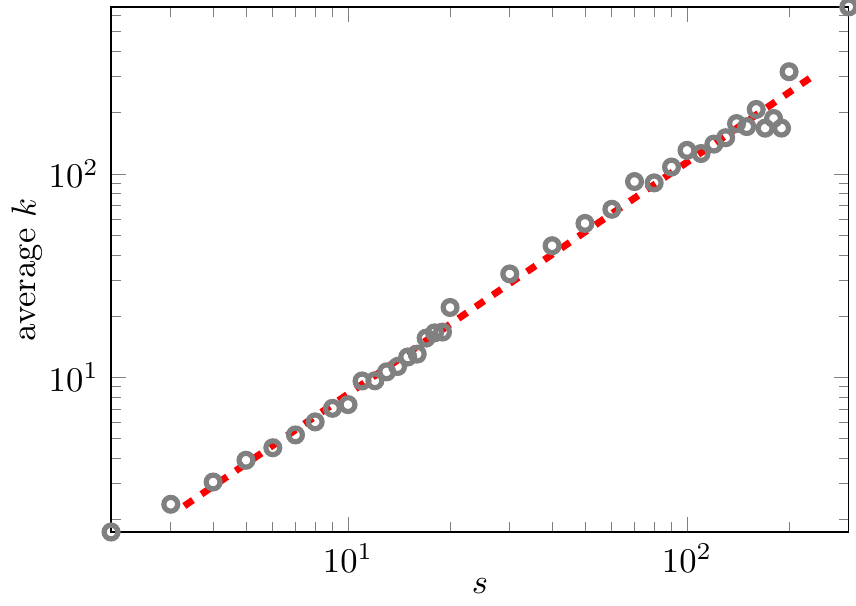}
		
		\label{fig:amaks}
	}	
	\hspace{0.2cm}
	\subfloat[]{
		\centering
		\includegraphics[width=0.35\textwidth]{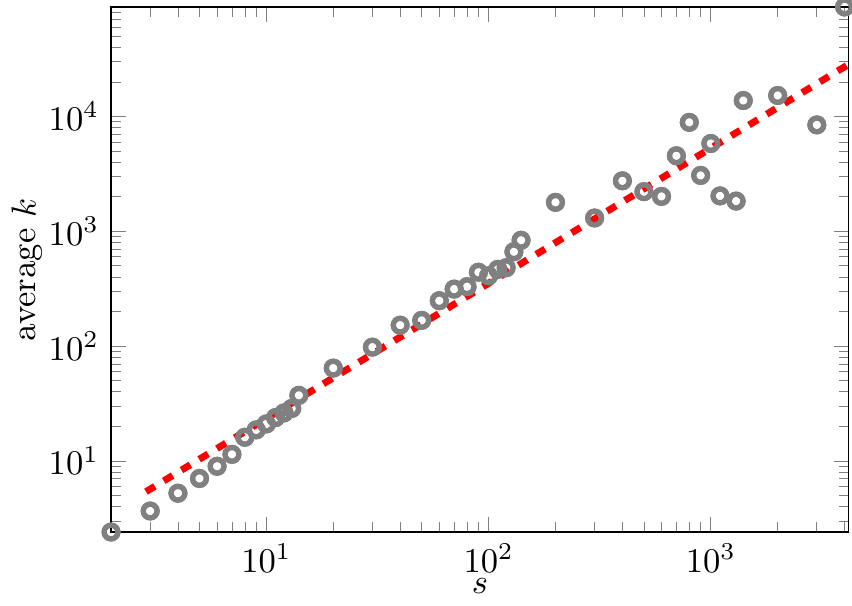}
		
		\label{fig:astrosk}
	}
	
	\subfloat[]{
		\centering
		\includegraphics[width=0.35\textwidth]{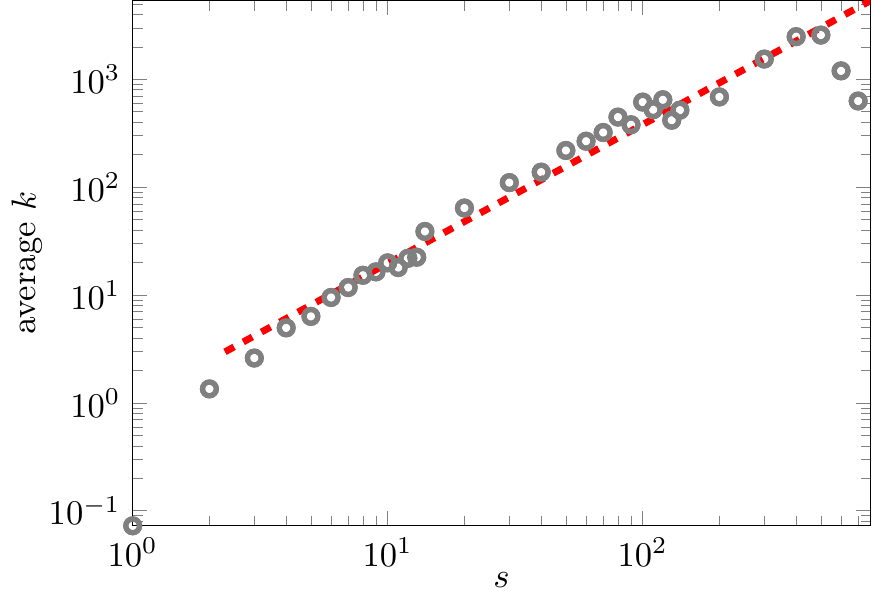}
		
		\label{fig:kscond}
	}
	\hspace{0.2cm}
	\subfloat[]{
		\centering
		\includegraphics[width=0.35\textwidth]{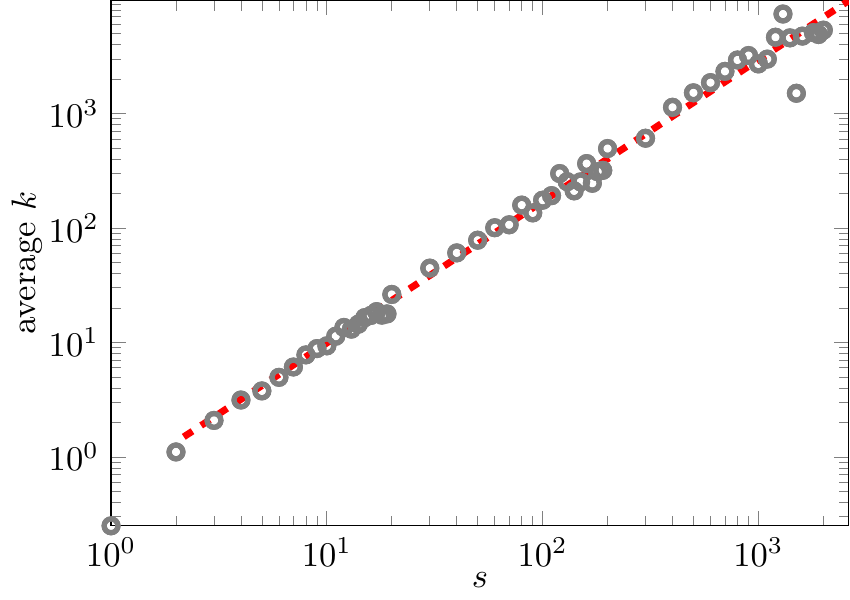}
		
		\label{fig:ksgoogle}
	}
	\caption{The outside degree of a community $k$ follows a power-law relation with the community size $s$. a) \textsc{Amazon} co-purchasing network, b) \textsc{Gowalla} social network, c) English word relations, d) \textsc{Google} web graph.}
	\label{fig:ks}	
\end{figure*}

We next test assumption~\eqref{eq:K}.
Interestingly, Figure~\ref{fig:ks} shows a power-law relationship between $k$ and $s$, of the form $k\approx s^\beta$. If $k\leq Ks^2$ would hold, then $\beta\leq 2$, whereas the more strict assumption~\eqref{eq:Kstrict} would imply $\beta<1$. Table~\ref{tab:alphabeta} shows that the example data sets all have $1<\beta<2$. Therefore, the more strict assumption~\eqref{eq:Kstrict} does not hold, but~\eqref{eq:K} does hold. Thus, large communities have very large outside degrees.

\section{Conclusions and outlook}\label{sec6}
We have introduced the Hierarchical Configuration Model (HCM) as a random graph model that can describe both realistic degree distributions and an arbitrary community structure, while remaining analytically tractable in the large-graph limit. Our analysis of the HCM has revealed several properties. The condition for a giant component to emerge in the HCM  is completely determined by properties of the macroscopic configuration model at the level of communities and therefore not affected by the precise structure or size of communities. The size of the giant component, however, strongly depends on the joint probability distribution describing the size of a community and its outside degree.
Under bond percolation, communities may either increase or decrease the critical percolation value compared to a configuration model with the same degree sequence.

For the prototypical case of extremely dense communities, we show that a power-law degree distribution with exponent $\tau$ implies a power-law distribution for the community sizes with exponent $\gamma=\tau+1$. Real-world networks, however, rarely posses an extremely dense community structure~\cite{leskovec2009}.

Studying the HCM allows us to observe two previously unobserved power-law relations in several real-world networks.
The relation between the number of edges inside a community $e_\text{in}$ and the community sizes $s$ follows a power law of the form $e_{\text{in}}\propto s^{1+\alpha}$. The second power-law relation is between the number of edges going out of a community $k$ and the community sizes: $k\propto s^{\beta}$.
The data sets that were studied in this paper had $1<\beta<2$ and $\alpha<1$.
Combined, the two power-law relations improve our understanding of the community structure in the data sets. Large communities are not extremely densely connected, and have a large number of edges going out of the community per vertex. Smaller communities are dense, and vertices in the community have only a few edges going out of the community. Our intuitive picture of extremely densely connected communities thus only holds for the small communities in a network, the larger communities do not fit into this picture.
The observation that large communities are not extremely dense may be a consequence of not allowing for overlapping communities. In case of several overlapping communities, community detection algorithms may merge these communities into one large community. As a consequence, this large community will be far from extremely dense. In the case of overlapping communities, many networks still display a power-law community size distribution~\cite{palla2005}. It would be interesting to investigate the relation between the exponent of the degree distribution and the community size distribution when communities are allowed to overlap. Further research could also study how the denseness of the communities and the number of edges out of the communities are related to the community sizes in the case of overlapping communities.

Both power-law relations are observational, and therefore depend on the Infomap community detection algorithm. It is also possible to use other community detection algorithms to investigate these power-law relations. We found that when using the Louvain community detection algorithm~\cite{blondel2008}, the power-law relations still hold. The estimates for the exponents $\alpha$ and $\beta$ however did change. This can be explained by the fact that the Louvain method finds larger communities in general, which are therefore less dense.

The power-law exponent of the degree distribution $\tau$ is known to influence the behavior of various processes on random graphs, for example percolation or epidemic models. Furthermore, mean distances in random graphs are different for $\tau\in(2,3)$, or $\tau>3$~\cite{hofstad2005,hofstad2007}.
Our results suggest that for networks with a community structure it is not clear whether the behavior of these processes can be explained by $\tau$, or the exponent of the community degrees, this remains open for further research. The results on the power-law shift suggest that this may depend on the density of the communities, which is characterized by the exponent $\alpha$.

The HCM keeps all edges inside the communities, while rewiring the inter-community edges.
Instead of fixing the precise internal community structure, one could also randomize the edges inside communities as in a CM.
This model was introduced as the modular random graph~\cite{sah2014}, a random graph with a given degree distribution and modularity. The focus in~\cite{sah2014} is on the algorithmic construction of the modular random graph, not on the analytical properties of this model. The analytic study of the modular random graph is worthwhile to pursue. From the present work, it is clear that the analysis of the giant component remains the same as for the HCM, at least when the communities are likely to be connected, so the precise details of the internal community structure can be safely ignored. However, we have also seen that the internal community structure does become important when considering the critical percolation threshold, and in this case the analysis of the HCM does not carry over to the modular random graph.

\subsection*{Acknowledgement}
This work is supported by NWO TOP grant 613.001.451 and by the NWO Gravitation Networks grant 024.002.003.
The work of RvdH is further supported by the NWO VICI grant 639.033.806.  The work of JvL is further supported by an NWO TOP-GO grant and by an ERC Starting Grant.

\appendix
\section{Influence of communities on percolation}\label{sec:exampleperc}
In this section, we study two examples of community structures. The first example decreases the critical percolation value when comparing the HCM with the CM with the same degree distribution. The second example increases the critical percolation value when comparing the HCM with the CM.

\begin{figure}[tb]
\centering
\includegraphics[width=0.25\textwidth]{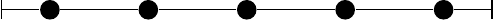}
\caption{A line community with $l=5$}
\label{fig:line}
\end{figure}

\begin{figure}[tb]
\centering
\includegraphics[width=0.38\textwidth]{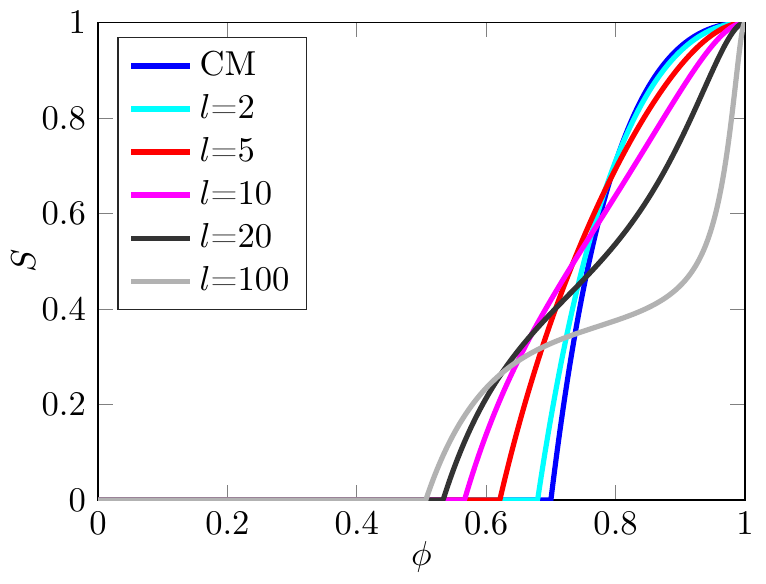}
\caption{Size of largest percolating cluster $S$ for HCM with line communities of length $l$, while the degree distribution remains the same. The line communities decrease the critical percolation value.}
\label{fig:linepi}
\end{figure}
As an example of a graph that decreases $\phi_c$, consider a network where with probability $\zeta$ a community is given by $H_1$: a path of $l$ vertices, with an outgoing half-edge at each end of the path as illustrated in Figure~\ref{fig:line}. With probability $1-\zeta$ the community is $H_2$: a vertex with three outgoing half-edges.
Then $\mean{k}=2\zeta+3(1-\zeta)=3-\zeta$ and $f(H_1,v,2,\phi)=\phi^{l-1}$ for all $v\in H_1$. In $H_2$ there is no percolation inside the community, hence $f(H_2,v,3,\phi)=1$. Using~\eqref{eq:pic} yields
	\begin{equation}
	\phi_c=\frac{3-\zeta}{ 2\zeta\phi_c^{l-1}+6(1-\zeta)},
	\end{equation}
hence $2\zeta\phi^{l}_c+6(1-\zeta)\phi_c-3+\zeta=0$.

Consider this version of the HCM with degree distribution $p_2=1-p_3=\frac23$. We keep the degree distribution the same, while the length $l$ of the line communities $H_1$ changes. Then if $l$ increases, $\zeta$ decreases to keep the degree distribution the same.
Using~\eqref{eq:uphi} and~\eqref{eq:Sphi}, we find the size of the giant component, as depicted in Figure~\ref{fig:linepi}. Note that $\phi_c$ decreases with $l$, this community structure `helps' the diffusion process. This can be explained by the fact that there will be fewer line communities if $l$ increases. Then most vertex communities are connected to one another, which decreases the value of $\phi_c$.
Interestingly, the size of the giant component in the HCM is non-convex in $\phi$. This is different than in the CM, where the percolation curves are typically convex.
The non-convex shapes can be explained intuitively. As the lines get longer, there are fewer and fewer of them, since the degree distribution remains the same. Hence, if $l$ is large, there are only a few very long lines. These lines have $\phi_c=1$. The other vertices are of degree 3, connected as the CM. Since there are only a few lines, most vertices of degree 3 will be paired to one another. The critical value for percolation on a CM with only vertices of degree 3 is $\frac12$. Therefore, for large $l$ we see the vertices of degree 3 appearing in the giant component as $\phi=0.5$, and the vertices in the line communities as $\phi=1$.

\begin{figure}[tb]
\centering
\includegraphics[width=0.15\textwidth]{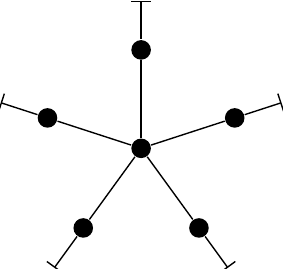}
\caption{A star-shaped community with $l=5$}
\label{fig:star}
\end{figure}

An example of a network that inhibits the diffusion process is a CM with intermediate vertices~\cite{litvak2013}, where every edge is replaced by two edges and a vertex in between them. This is the same as the HCM with star-shaped communities: one vertex that is connected to $l$ other vertices. Each of the $l$ other vertices has outside degree one (Figure~\ref{fig:star}). In this example, we consider a HCM where all communities have the same star-size $l$, so that $\mean{k}=l$.
After percolation, the connected component of an end point of the star can link to other outgoing edges only if the edge to the middle vertex is present. If this edge is present, the number of half-edges to which the vertex is connected is binomially distributed:
	\begin{equation}
	f(H,v,k,\phi)=\phi{{l-1}\choose{k-1}}\phi^{k-1}(1-\phi)^{l-k}\quad k\geq 2.
	\end{equation}
Hence by~\eqref{eq:pic},
	\begin{equation}
	\begin{aligned}[b]
	\phi_c&=\frac{l}{l\phi_c\sum_{k\geq 1}k\phi_c^k(1-\phi_c)^{l-k-1}{l-1\choose k}}\\
	&=\frac{1}{(l-1)\phi_c^2},
	\end{aligned}
	\end{equation}
so that $\phi_c=(l-1)^{-1/3}$.

\begin{figure}[htb]
\centering
\includegraphics[width=0.38\textwidth]{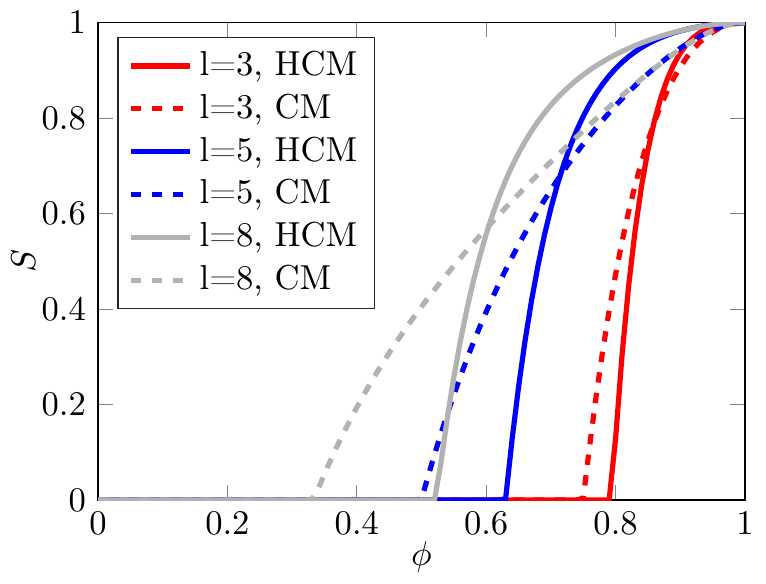}
\caption{Size of largest percolating cluster $S$ for the HCM with star communities of size $l$, compared to the CM. The star communities increase the the critical percolation value.}
\label{fig:starpi}
\end{figure}

We next consider a CM with the same degree distribution.
Figure~\ref{fig:starpi} shows the size of the giant component for different values of $l$ for both the HCM and the CM. The HCM with star communities has a higher critical percolation value than the corresponding CM. Intuitively, this can be explained by the fact that all vertices with a high degree are `hidden' behind vertices of degree 2, whereas in the CM, vertices with degree $l$ may be connected to one another. However, as $\phi$ increases, at some point, the star communities make the giant component larger.

Combined with the previous example, we see that adding communities may lead to a higher critical percolation value or a lower one. Furthermore, the size of the giant component may be smaller or larger after adding communities.

\bibliographystyle{abbrv}
\bibliography{../references}

\begin{thebibliography}{10}

\bibitem{albert1999}
R.~Albert, H.~Jeong, and A.-L. Barab{\'a}si.
\newblock Internet: Diameter of the world-wide web.
\newblock {\em Nature}, 401(6749):130--131, 1999.

\bibitem{ball2009}
F.~Ball, D.~Sirl, and P.~Trapman.
\newblock Threshold behaviour and final outcome of an epidemic on a random
  network with household structure.
\newblock {\em Adv. in Appl. Probab.}, 41(3):765--796, 09 2009.

\bibitem{ball2010}
F.~Ball, D.~Sirl, and P.~Trapman.
\newblock Analysis of a stochastic {SIR} epidemic on a random network
  incorporating household structure.
\newblock {\em Mathematical Biosciences}, 224(2):53 -- 73, 2010.

\bibitem{Barabasi2000}
A.-L. Barab{\'a}si, R.~Albert, and H.~Jeong.
\newblock Scale-free characteristics of random networks: the topology of the
  world-wide web.
\newblock {\em Physica A: Statistical Mechanics and its Applications},
  281(1–4):69 -- 77, 2000.

\bibitem{bhamidi2010}
S.~Bhamidi, R.~v.~d. {\swap{Hofstad}{van der }}, and G.~Hooghiemstra.
\newblock First passage percolation on random graphs with finite mean degrees.
\newblock {\em The Annals of Applied Probability}, 20(5):pp. 1907--1965, 2010.

\bibitem{blondel2008}
V.~D. Blondel, J.-L. Guillaume, R.~Lambiotte, and E.~Lefebvre.
\newblock Fast unfolding of communities in large networks.
\newblock {\em Journal of statistical mechanics: theory and experiment},
  2008(10):P10008, 2008.

\bibitem{boguna2004}
M.~Bogu\~n\'a, R.~Pastor-Satorras, A.~D\'{\i}az-Guilera, and A.~Arenas.
\newblock Models of social networks based on social distance attachment.
\newblock {\em Phys. Rev. E}, 70:056122, Nov 2004.

\bibitem{bollobas1980}
B.~Bollob{\'a}s.
\newblock A probabilistic proof of an asymptotic formula for the number of
  labelled regular graphs.
\newblock {\em European Journal of Combinatorics}, 1(4):311--316, 1980.

\bibitem{callaway2000}
D.~S. Callaway, M.~E.~J. Newman, S.~H. Strogatz, and D.~J. Watts.
\newblock Network robustness and fragility: Percolation on random graphs.
\newblock {\em Phys. Rev. Lett.}, 85(25):5468, 2000.

\bibitem{cho2011}
E.~Cho, S.~A. Myers, and J.~Leskovec.
\newblock Friendship and mobility: user movement in location-based social
  networks.
\newblock In {\em Proceedings of the 17th ACM SIGKDD international conference
  on Knowledge discovery and data mining}, pages 1082--1090. ACM, 2011.

\bibitem{clauset2004}
A.~Clauset, M.~E.~J. Newman, and C.~Moore.
\newblock Finding community structure in very large networks.
\newblock {\em Phys. Rev. E}, 70:066111, Dec 2004.

\bibitem{clauset2009}
A.~Clauset, C.~R. Shalizi, and M.~E.~J. Newman.
\newblock Power-law distributions in empirical data.
\newblock {\em SIAM Review}, 51(4):661--703, 2009.

\bibitem{cohen2001}
R.~Cohen, K.~Erez, D.~Ben-Avraham, and S.~Havlin.
\newblock Breakdown of the internet under intentional attack.
\newblock {\em Phys. Rev. Lett.}, 86:3682--3685, Apr 2001.

\bibitem{coupechoux2014}
E.~Coupechoux and M.~Lelarge.
\newblock How clustering affects epidemics in random networks.
\newblock {\em Adv. in Appl. Probab.}, 46(4):985--1008, 12 2014.

\bibitem{faloutsos1999}
M.~Faloutsos, P.~Faloutsos, and C.~Faloutsos.
\newblock On power-law relationships of the internet topology.
\newblock In {\em ACM SIGCOMM Computer Communication Review}, volume~29, pages
  251--262. ACM, 1999.

\bibitem{fortunato2010}
S.~Fortunato.
\newblock Community detection in graphs.
\newblock {\em Physics Reports}, 486(3):75--174, 2010.

\bibitem{guimera2005}
R.~Guimera and L.~A.~N. Amaral.
\newblock Functional cartography of complex metabolic networks.
\newblock {\em Nature}, 433(7028):895--900, 2005.

\bibitem{guimera2003}
R.~Guimer\`a, L.~Danon, A.~D\'{\i}az-Guilera, F.~Giralt, and A.~Arenas.
\newblock Self-similar community structure in a network of human interactions.
\newblock {\em Phys. Rev. E}, 68:065103, Dec 2003.

\bibitem{hofstad2005}
R.~{\swap{Hofstad}{van der }}, G.~Hooghiemstra, and P.~{\swap{Mieghem}{Van }}.
\newblock Distances in random graphs with finite variance degrees.
\newblock {\em Random Structures \& Algorithms}, 27(1):76--123, 2005.

\bibitem{hofstad2007}
R.~{\swap{Hofstad}{van der }}, G.~Hooghiemstra, and D.~Znamenski.
\newblock Distances in random graphs with finite mean and infinite variance
  degrees.
\newblock {\em Electron. J. Probab.}, 12:no. 25, 703--766, 2007.

\bibitem{hofstad2015}
R.~{\swap{Hofstad}{van der }}, J.~S.~H. {\swap{Leeuwaarden}{van }}, and
  C.~Stegehuis.
\newblock Hierarchical configuration model.
\newblock Preprint 2015.

\bibitem{janson2008}
S.~Janson.
\newblock On percolation in random graphs with given vertex degrees.
\newblock {\em Electron. Journal of Probability}, 14:86--118, 2009.

\bibitem{karrer2010}
B.~Karrer and M.~E.~J. Newman.
\newblock Random graphs containing arbitrary distributions of subgraphs.
\newblock {\em Phys. Rev. E}, 82:066118, Dec 2010.

\bibitem{lancichinetti2009}
A.~Lancichinetti and S.~Fortunato.
\newblock Community detection algorithms: a comparative analysis.
\newblock {\em Physical review E}, 80(5):056117, 2009.

\bibitem{leskovec2009}
J.~Leskovec, K.~J. Lang, A.~Dasgupta, and M.~W. Mahoney.
\newblock Community structure in large networks: Natural cluster sizes and the
  absence of large well-defined clusters.
\newblock {\em Internet Mathematics}, 6(1):29--123, 2009.

\bibitem{litvak2013}
N.~Litvak and R.~{\swap{Hofstad}{van der }}.
\newblock Uncovering disassortativity in large scale-free networks.
\newblock {\em Phys. Rev. E}, 87:022801, Feb 2013.

\bibitem{miller1998}
G.~Miller and C.~Fellbaum.
\newblock Wordnet: An electronic lexical database, 1998.

\bibitem{molloy1995}
M.~Molloy and B.~Reed.
\newblock A critical point for random graphs with a given degree sequence.
\newblock {\em Random Structures \& Algorithms}, 6(2-3):161--180, 1995.

\bibitem{newman2003book}
M.~E.~J. Newman.
\newblock The structure and function of complex networks.
\newblock {\em SIAM Review}, 45(2):167--256, 2003.

\bibitem{newman2009}
M.~E.~J. {Newman}.
\newblock Random graphs with clustering.
\newblock {\em Phys. Rev. Lett.}, 103(5):058701, July 2009.

\bibitem{newman2010}
M.~E.~J. Newman.
\newblock {\em Networks: an introduction}.
\newblock Oxford University Press, 2010.

\bibitem{newman2002b}
M.~E.~J. Newman, S.~Forrest, and J.~Balthrop.
\newblock Email networks and the spread of computer viruses.
\newblock {\em Phys. Rev. E}, 66:035101, Sep 2002.

\bibitem{newman2001}
M.~E.~J. Newman, S.~H. Strogatz, and D.~J. Watts.
\newblock Random graphs with arbitrary degree distributions and their
  applications.
\newblock {\em Phys. Rev. E}, 64(2):026118, 2001.

\bibitem{palla2005}
G.~Palla, I.~Der{\'e}nyi, I.~Farkas, and T.~Vicsek.
\newblock Uncovering the overlapping community structure of complex networks in
  nature and society.
\newblock {\em Nature}, 435(7043):814--818, 2005.

\bibitem{radicchi2004}
F.~Radicchi, C.~Castellano, F.~Cecconi, V.~Loreto, and D.~Parisi.
\newblock Defining and identifying communities in networks.
\newblock {\em Proceedings of the National Academy of Sciences of the United
  States of America}, 101(9):2658--2663, 2004.

\bibitem{Rosvall2008}
M.~Rosvall and C.~T. Bergstrom.
\newblock Maps of random walks on complex networks reveal community structure.
\newblock {\em Proceedings of the National Academy of Sciences},
  105(4):1118--1123, 2008.

\bibitem{sah2014}
P.~Sah, L.~O. Singh, A.~Clauset, and S.~Bansal.
\newblock Exploring community structure in biological networks with random
  graphs.
\newblock {\em BMC Bioinformatics}, 15(1):220, 2014.

\bibitem{trapman2007}
P.~Trapman.
\newblock On analytical approaches to epidemics on networks.
\newblock {\em Theoretical Population Biology}, 71(2):160 -- 173, 2007.

\bibitem{vazquez2002}
A.~V{\'a}zquez, R.~Pastor-Satorras, and A.~Vespignani.
\newblock Large-scale topological and dynamical properties of the internet.
\newblock {\em Phys. Rev. E}, 65(6):066130, 2002.

\bibitem{yang2015}
J.~Yang and J.~Leskovec.
\newblock Defining and evaluating network communities based on ground-truth.
\newblock {\em Knowledge and Information Systems}, 42(1):181--213, 2015.

\end{thebibliography}
\end{document}